\documentclass[prl,twocolumn,showpacs,superscriptaddress]{revtex4-1}

\usepackage{graphicx}
\usepackage[usenames,dvipsnames]{color} 
\usepackage{amsmath} 
\usepackage{amssymb}
\usepackage{soul}

\newcommand{\dg}{^\dagger}
\def\be#1\ee{\begin{equation}#1\end{equation}}
\def\ba#1\ea{\begin{align}#1\end{align}}
\def\bg#1\eg{\begin{gather}#1\end{gather}}
\def\t{\text}
\def\n{\nonumber}

\newcommand{\B}[1]{\mathbf{#1}}
\newcommand{\bs}[1]{\boldsymbol{#1}}
\def\eq#1{(\ref{eq:#1})}

\newcommand{\sfrac}[2]{\mbox{$\frac{#1}{#2}$}}

\begin{document}

\title{Quantum non-demolition measurement of microwave photons using engineered quadratic interactions}

\author{Chunqing Deng}
\affiliation{Institute for Quantum Computing and the Department of Physics and Astronomy, University of Waterloo, Waterloo, ON, Canada N2L 5G7}
\author{J. M. Gambetta}
\affiliation{Institute for Quantum Computing and the Department of Physics and Astronomy, University of Waterloo, Waterloo, ON, Canada N2L 5G7}
\author{A. Lupa\c scu}
\affiliation{Institute for Quantum Computing and the Department of Physics and Astronomy, University of Waterloo, Waterloo, ON, Canada N2L 5G7}

\date{ \today}

\begin{abstract}

We present a quantum electrical circuit with Josephson junctions formed of two anharmonic oscillators coupled with an interaction $g\gamma_{1}^{2}\gamma_{2}^{2}$ where $\gamma_{1}$ and $\gamma_{2}$ are position-like coordinates. This type of coupling allows the quantum non-demolition measurement of the energy of one oscillator by monitoring the frequency of the second oscillator. Despite the fundamental tradeoff between the coupling strength $g$ and maximum photon storage capacity of the oscillators, it is possible to achieve high fidelity detection of up to 10 photons over time scale of the order of microseconds. We discuss the possibility of observing quantum jumps in the number of photons and related applications.

\end{abstract}
\pacs{85.25.Cp
, 42.50.Dv
, 03.65.Ta
, 85.25.Dq
}
\maketitle

The quantum measurement of harmonic oscillators is a generic problem in quantum mechanics. One context in which this problem was analyzed is the detection of gravitational waves~\cite{thorne_1980_GravitationalWaves}. Another domain of application is quantum optics, where the measurement of the number of photons in single or multiple modes of electromagnetic fields is an essential tool for quantum field state tomography~\cite{deleglise_2008_CavityFieldsAndDecoherence,Hofheinz2009}, linear quantum computation \cite{Knill2001}, and quantum communication \cite{Gisin2007}. Most measurement schemes are hampered by either limited efficiency or large backaction. An ideal measurement of an oscillator in the basis of Fock states can be achieved by employing a quantum non-demolition (QND) strategy~\cite{braginsky_1992_1}. In contrast to non-ideal measurements, QND schemes offers interesting prospects for preparation of Fock states on demand and application to detection of weak forces \cite{caves_1980_WeakForce1}.

QND detection of Fock states requires an interaction between the observed oscillator and the employed detector which depends on the position and/or momentum of the former in a non-linear way. Finding a suitable non-linear interaction is a difficult task. Up to date, this has only been achieved in a few experiments. Examples include the measurement of an electron in a Penning trap~\cite{peil_1999_QNDcyclotron}, the measurement of microwave photons in a Fabry-Perot superconducting resonator~\cite{guerlin_2007_QNDStateCollapse}, and the measurement of microwave photons in an on-chip superconducting cavity~\cite{johnson_2010_QNDSinglePhotons}. Furthermore, other QND measurement schemes based on nonlinear interactions have been proposed for nanomechanical systems~\cite{santamore_2004_QNDFockresonators} and for superconducting resonators~\cite{PhysRevB.82.014512}.

In this letter we propose a method for the QND measurement of Fock states of a simple lumped electrical oscillator formed by a direct-current superconducting quantum interference device (DC-SQUID) and a capacitor. The Fock states are photon number states stored in this microwave resonator. We show that it is possible to design an interaction between the observed resonator and a second resonator which is quadratic in position-like coordinates. Combined with a proper choice of resonant frequencies of the two resonators (\emph{i.e.} sufficiently large detuning) this interaction can be used to measure the number of photons in the resonator of interest by monitoring the frequency of the second resonator. This method for QND detection of microwave photons is different to the experiment in~\cite{johnson_2010_QNDSinglePhotons}, which was based on using the dispersive regime of the Jaynes Cummings Hamiltonian~\cite{Gambetta2006}, and the proposal in~\cite{PhysRevB.82.014512}, which uses Josephson junctions to couple different modes of a single resonator. Our method has the advantage of very strong achievable coupling resulting in a large signal to noise ratio (SNR) for detection of up to 10 photons in the resonator. We perform a detailed analysis of the optimization of the non-linear coupling used for measurement and the tradeoff between the strength of coupling and anharmonicity of each circuit.

\begin{figure}[!]
\includegraphics[width=3.4in]{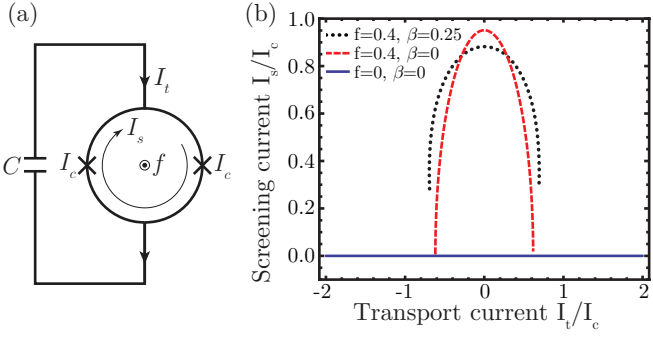}
\caption{\label{fig1} (a) DC-SQUID based resonator: the cross symbols represent Josephson junctions. (b) (Color online) Screening current $I_s$ versus transport current $I_t$ for $f=0$, $\beta=0$ (continuous blue line), $f=0.4$, $\beta=0$ (dashed red line), and $f=0.4$, $\beta=0.2$ (dotted black line ), see text for details.}
\end{figure}

The electrical resonator that we consider has the simple structure shown in Fig.~\ref{fig1}a. It can be seen as an electrical LC circuit formed of the capacitance $C$ and a SQUID acting as an effective inductor. The SQUID is a two-terminal superconducting device formed of a superconducting loop interrupted by two Josephson junctions. A SQUID can be used as a sensitive magnetometer, based on the property that its critical current depends on the magnetic flux $\Phi$ enclosed in its ring with a period equal to the flux quantum $\Phi_{0}=h/2e$~\cite{barone_1982_1}. For a current in the SQUID below this critical current, the SQUID behaves as an inductor, with an inductance that depends on the magnetic flux. Therefore, the circuit in Fig.~\ref{fig1}a can be seen as an LC resonator, with a resonance frequency dependent on the magnetic flux enclosed in the ring. As such, this system was used for the readout of a flux qubit~\cite{lupascu_2004_1}.

At non-integer values of the applied magnetic flux, a finite \emph{screening} current $I_{s}$ (circulating in the ring of the SQUID) arises. This is illustrated in Fig.~\ref{fig1}b for different values of the applied flux $f=\Phi/\Phi_0$ and ratio $\beta=L/L_J$ of the geometric inductance of the SQUID ring, $L$, to the Josephson inductance of each junction $L_J$, with $L_{J}=\varphi_{0}/ I_c$ ($\varphi_{0}=\Phi_0 /2\pi$). For a symmetric SQUID this screening current is an even function of the \emph{transport} current $I_{t}$, and in particular it has a strong quadratic component. Therefore, different energy eigenstates in the SQUID, characterized by different transport current \emph{quantum fluctuations}, will result in different average values of the screening current. A second SQUID based LC-resonator, acting as a flux detector, can then be used to measure the screening current and therefore implement a measurement of the energy, or equivalently of the photon Fock states, of the first SQUID. This is the basis of the principle of detection. In the following we present a full quantum mechanical calculation necessary to validate and understand the limits of this qualitative argument.

\begin{figure}[!]
\includegraphics[width=3.4in]{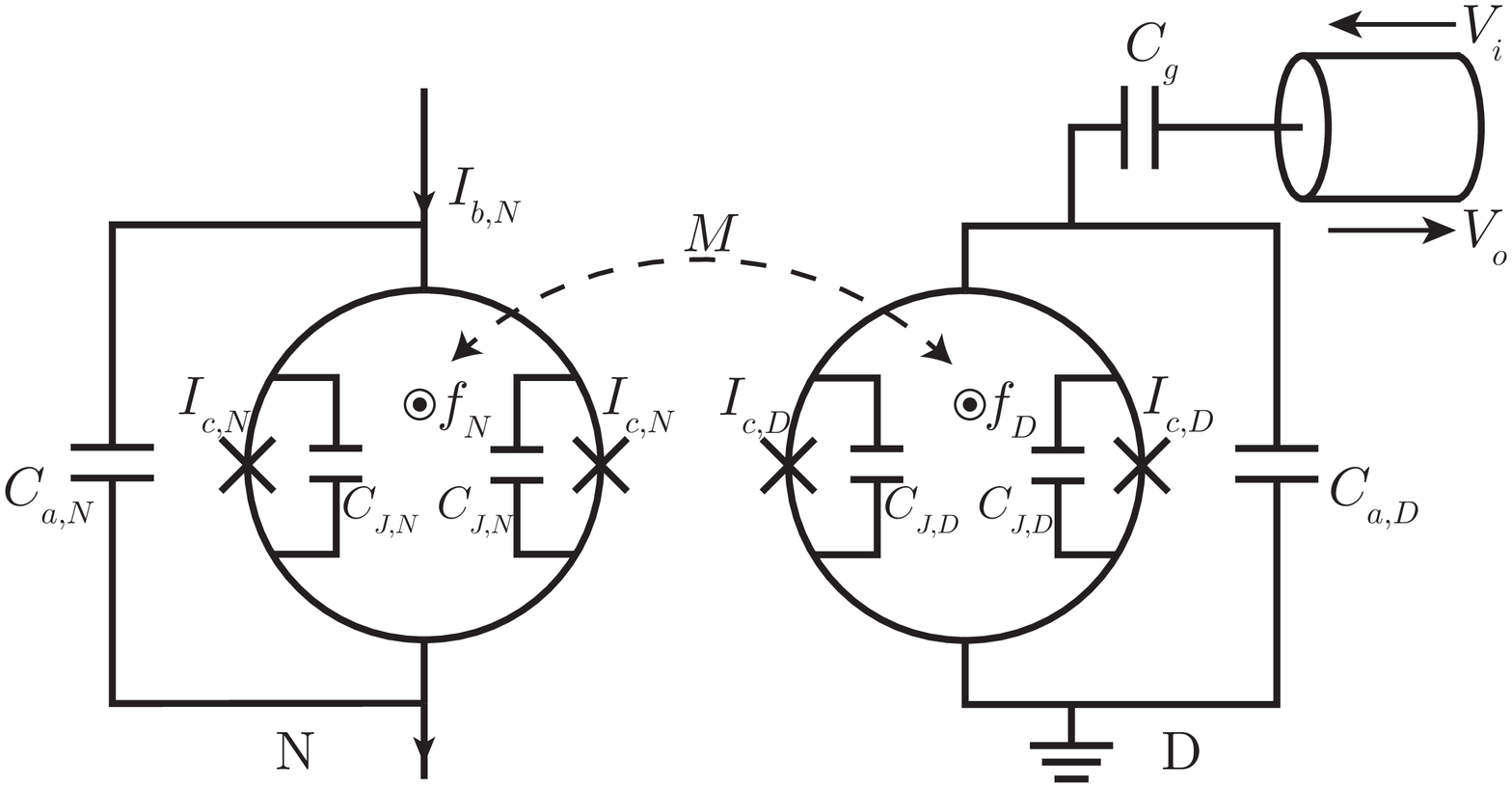}
\caption{\label{fig2} Diagram of the measurement scheme (see text for details). The presence of $n$ photons in the \emph{N} resonator causes an $n$-dependent shift of the frequency of the \emph{D} resonator, which is monitored in a reflection measurement.}
\end{figure}

Fig.~\ref{fig2} shows the full circuit details used to model the measured (\emph{N}) and detection (\emph{D}) resonators. The two SQUIDS are symmetric, with junctions critical currents $I_{c,\alpha}$, intrinsic junction capacitances $C_{J,\alpha}$, inductance $L_{\alpha}$ in each arm (not shown), and magnetic flux $f_{\alpha}$ applied to each loop using an external source (not shown), with $\alpha=N,D$. Each SQUID is shunted by a capacitor $C_{a,\alpha}$ with a much larger capacitance than the junction capacitors. The two SQUIDS are coupled through their mutual inductance $M$. A bias current source source $I_{b,N}(t)$ is used to control the state of the observed resonator. The detection resonator is coupled using a coupling capacitor $C_{g}$ to a transmission line of characteristic impedance $Z_{0}$. A wave of amplitude $V_{i}$ and frequency $\bar\omega_{D}$ is sent to the detector resonator; the reflected wave amplitude, $V_{o}$, is measured and the result is used to infer the state of the \emph{N} resonator.

We applied the rules of circuit quantization~\cite{burkard_2004_1} to derive the quantum model of the complete circuit shown in Fig.~\ref{fig2}. The two resonators can be described by six degrees of freedom: the gauge invariant phases $\gamma_{i,\alpha}$ over the Josephson junctions ($i=1,2$ labels the two Josephson junctions in each SQUID) and the phases $\gamma_{a,\alpha}$ (defined as the flux of circuit theory divided by the reduced flux quantum $\varphi_{0}=\Phi_{0}/2\pi$~\cite{burkard_2004_1}) over the shunt capacitors $C_{a,\alpha}$. We use~\cite{burkard_2004_1} to derive the Hamiltonian of the complete system:
\be
\mathcal{H}=\sum_{\alpha=N,D}T_{\alpha}+\sum_{\alpha=N,D}U_{J,\alpha}+U_L+H_{\t{ctr},N}+H_{\t{meas},D} \label{eq:H1}.
\ee
$T_{\alpha}=\frac{1}{2}p_{s,\alpha}^2/(\varphi_0^2 2C_{J,\alpha})+\frac{1}{2}p_{t,\alpha}^2/(\varphi_0^2 2C_{J,\alpha})+\frac{1}{2}p_{a,\alpha}^2/(\varphi_0^2 \bar{C}_{a,\alpha})$ is the sum of the electric energies in the capacitors in resonator $\alpha$ with $p_{s,\alpha}$, $p_{t,\alpha}$, and $p_{a,\alpha}$ the momenta canonically conjugate to the variables $\gamma_{s,\alpha}=(\gamma_{1,\alpha}-\gamma_{2,\alpha})/2$, $\gamma_{t,\alpha}=(\gamma_{1,\alpha}+\gamma_{2,\alpha})/2$, and $\gamma_{a,\alpha}$ respectively; $\bar{C}_{a,N}=C_{a,N}$ and $\bar{C}_{a,D}=C_{a,D}+C_{g}$. The Josephson energy in each SQUID is given by $U_{J,\alpha}=-2\varphi_0 I_{c,\alpha}\cos\gamma_{t,\alpha} \cos\gamma_{s,\alpha}$. The total magnetic energy stored in the two circuits is $U_L=\sum_{\alpha,\beta}2\varphi_0^2L_{\alpha\beta}^{-1}(\gamma_{s,\alpha}+\pi f_\alpha)(\gamma_{s,\beta}+\pi f_\beta)+\sum_{\alpha,\beta}2\varphi_0^2 {L'}_{\alpha\beta}^{-1}(\gamma_{t,\alpha}-\gamma_{a,\alpha})(\gamma_{t,\beta}-\gamma_{a,\beta})$, with the inductance matrices $\B{L}=\bigl( \begin{smallmatrix}
  L_{N} & M\\
  M & L_{D}
\end{smallmatrix} \bigr) $ and $\B{L}'=\bigl( \begin{smallmatrix}
  L_{N}+4L_{a,N} & -M\\
  -M & L_{D}+4L_{a,D}
\end{smallmatrix} \bigr) $. $L_{a,\alpha}$ is a small unavoidable stray inductance associated with $C_{a,\alpha}$. The \emph{N} resonator is controlled through $H_{\t{ctr},N}=-\varphi_0I_{b,N}(t)\gamma_{a,N}$, by using $I_{b,N}(t)$. Finally, the probing of the \emph{D} resonator is done by coupling it to a transmission line: $H_{\t{meas},D}=-C_gp_{a,D}V_{i}/\bar C_{a,D}\varphi_0$.

We choose $C_{a,\alpha}$ to be significantly larger than $C_{J,\alpha}$. As a consequence, the degrees of freedom $\gamma_{s,\alpha}$ and $\gamma_{t,\alpha}$ have a much smaller effective mass and therefore they can be eliminated from the problem using the Born-Oppenheimer approximation. Assuming that these fast degrees of freedom remain in the ground state (which is justified in view of the experimental parameters we choose), we determine the effective potential energy of the slow degrees of freedom by minimizing the potential energy with respect to $\gamma_{s,\alpha}$ and $\gamma_{t,\alpha}$. The potential energy terms $U_{J,\alpha}$ and $U_L$ in Eq.~\eq{H1} are replaced by $U'_{J,\alpha}$ and $U'_L$, respectively, which now only depend on $\gamma_{a,\alpha}$. We calculate the effective energy in the first order approximation with respect to $\beta_{\alpha}\cos{\pi f_{\alpha}}$, with $\beta_{\alpha}=L_{\alpha\alpha}/L_{J,\alpha}$ and we find:
\be
 U'_{J,\alpha}=-\frac{2\varphi_0^2\cos\pi f_\alpha \cos \gamma_{a,\alpha}}{L_{J,\alpha}}
\ee
and
\ba
U'_L=&-\frac{\varphi_0^2}{2}(\B{L}_J^{-1}\sin\pi{\B f}\cos\bs{\gamma}_a)^T{\B L}({\B{L}}_J^{-1}\sin\pi{\B f}\cos\bs{\gamma}_a) \n \\
-&\frac{\varphi_0^2}{2}({\B{L}}_J^{-1}\cos\pi{\B f}\sin\bs{\gamma}_a)^T{\B L}'({\B{L}}_J^{-1}\cos\pi{\B f}\sin\bs{\gamma}_a).
\ea
${\B{L}}_J=\bigl( \begin{smallmatrix}
  L_{J,N} & 0\\
  0 & L_{J,D}
\end{smallmatrix} \bigr)$ and ${\B f}=\bigl( \begin{smallmatrix}
 f_N & 0\\
  0 & f_D
\end{smallmatrix} \bigr)$ are the Josephson inductance matrix and bias flux matrix respectively, where we have defined the Josephson inductance $L_{J,\alpha}=\varphi_0/I_{c,\alpha}$. $\bs{\gamma}_a=\bigl( \begin{smallmatrix} \gamma_{a,N} \\ \gamma_{a,D} \end{smallmatrix} \bigr)$ is a column vector containing variables $\gamma_{a,\alpha}$. We define $\cos \bs{\gamma}_a=\bigl( \begin{smallmatrix} \cos\gamma_{a,N} \\ \cos\gamma_{a,D} \end{smallmatrix} \bigr)$ and $\sin \bs{\gamma}_a=\bigl( \begin{smallmatrix} \sin\gamma_{a,N} \\ \sin\gamma_{a,D} \end{smallmatrix} \bigr)$.

After applying the rotating wave approximation, the Hamiltonian of the coupled system has the form
\ba
\mathcal{H}_{\t{eff}}'/\hbar=&[\omega_D+A_{D}(a\dg a-1)]a\dg a+[\omega_N+A_{N}(b\dg b-1)]b\dg b \n \\
&+J(a\dg b+ab\dg)+2\varepsilon_D(t)(a\dg+a)+ga\dg ab\dg b \label{eq:Hh}
\ea
where $a$ and $b$ are the annihilation operators for the \emph{D} and \emph{N} resonators respectively, $\varepsilon_D(t)=\bar\varepsilon_D\cos{\bar\omega_{D}t}$, with $\bar\varepsilon_D$ the driving amplitude, and we have left out the $H_{\mathrm{ctr},N}$ in Eq.~\eqref{eq:H1}. The resonance frequencies are given by $\omega_\alpha=\sqrt{p_\alpha/L_{J,\alpha} \bar C_{a,\alpha}}$, with $p_\alpha=2\cos\pi f_\alpha+\sum_\beta \frac{L_{\alpha\beta}}{L_{J,\beta}}\sin \pi f_\alpha \sin\pi f_\beta-\frac{L'_{\alpha\alpha}}{L_{J,\alpha}}\cos^2 \pi f_\alpha$. The other parameters in the model are given by
\bg
J=\frac{k\sqrt{\omega_D\omega_N}\sqrt{\beta_D\beta_N}\cos\pi f_D \cos\pi f_N}{2\sqrt{p_N p_D}} \label{eq:J}\\
g=-\frac{k\sqrt{\omega_D\omega_N}\sqrt{\beta_D\beta_N}\sin\pi f_D\sin \pi f_N} {4\sqrt{n_{\mathrm{max},N} n_{\mathrm{max},D}}\sqrt{p_N p_D}} \label{eq:g}\\
A_\alpha=\frac{\omega_\alpha}{4n_{\mathrm{max},\alpha} p_\alpha}\bigl( \frac{1}{3}\cos\pi f_\alpha+\frac{1}{2}\beta_\alpha \sin^2 \pi f_\alpha \bigr)
\eg
with $k=M/\sqrt{L_D L_N}$ the inductive coupling constant ($|k|\leq 1$), and $n_{\mathrm{max},\alpha}={\omega_\alpha R_{Q} \bar{C}_{a,\alpha}/8\pi}$ with $R_Q=h/e^2$ the quantum resistance. The maximum number of photons, $n_{\mathrm{max},\alpha}$, is the number of photons corresponding to a energy equal to approximately the modulation height of the potential energy for each SQUID; this is the maximum photon storage capacity for each resonator. Equation~\eqref{eq:g} shows that the optimal value of $g$ is at best $\sqrt{\omega_D\omega_N}/\sqrt{n_{\mathrm{max},N} n_{\mathrm{max},D}}$, multiplied by a factor of the order of the unity. With given resonant frequencies (fixed values of $\omega_D$ and $\omega_N$) $g$ can be increased at the expense of $n_{\mathrm{max},D}$, however this leaves the fidelity of measurement unchanged (see discussion below). An augmentation of $g$ is possible by reducing $n_{\mathrm{max},N}$.

To implement the measurement of the \emph{N} resonator, the \emph{D} resonator is continuously monitored through the coupled transmission line. We introduce the decay rate of the resonator $\kappa=\frac{1}{C_{a,D}}\mathfrak{Re}[1/(\frac{1}{j\omega C_g}+Z_0)]$, where $Z_0$ is the characteristic impedance of the transmission line. This decay rate is tunable by merely changing the coupling capacitance $C_g$.
The system is described by the master equation
\begin{equation}\dot R=-i[H,R]+\kappa(a R a^\dagger -\sfrac{1}{2}a^\dagger a R -\sfrac{1}{2} R a^\dagger a ),\label{eq:master}\end{equation} where $R$ is the quantum state of the combined detector and system. We assume that that we can neglect the self-anharmonicity of the D resonator and that $|\omega_D-\omega_N|\gg|J|$ so that the linear interaction can be ignored (it only shifts the frequencies of the resonators). We then follow the procedure outlined in Ref.~\cite{Gambetta_2008}: the reduced state $\rho=\mathrm{Tr}[R]=\sum_{mn}\rho_{mn}|{m}\rangle\langle{n}|$ of the N resonator is found to be
 $\rho_{mn}=\rho_{mn}(0)\exp[-i \int_0^t H^{mn}dt' -\int_0^t\Gamma_d^{mn}dt']$ where
\begin{equation}\begin{split}
H^{mn}&= H_b^{mn}+g(m-n) \mathrm{Re}[\alpha_n\alpha_m^*]\\
\Gamma_d^{mn}&=\frac{\kappa}{2}|\alpha_m-\alpha_n|^2+\frac{1}{2}d_t|\alpha_m-\alpha_n|^2
\end{split}
\label{eq:rates}\end{equation} with $H^{mn}_b=(\omega_N-A_N)(m-n)+A_N(m^2-n^2)$ and $\dot \alpha_n = -i\bar\varepsilon_D -i(\Delta+gn)\alpha_n-\kappa \alpha_n/2$. Here $\Delta=\omega_D-\bar\omega_D$ is the detuning between the driving and resonance frequency. In~\eqref{eq:rates} the first equation represents the Hamiltonian evolution of the \emph{N} resonator which comprises of both bare evolution (first term) and the ac-Stark shift induced by the detection resonator. The second equation represents dephasing: it comprises of a term proportional to the information extracted by the detector (see below) and  a second term which can be both positive and negative as it represents information flowing between the detector and system that has not been extracted by the monitoring.

In order to quantify the detection efficiency we consider the problem of distinguishing a $n$ from a $n+1$ photon state in the \emph{N} resonator. In this case the measurement rate is
\be
\Gamma_m^{n,n+1}=\kappa|\alpha_{n+1}-\alpha_n|^2.
\ee
This can be interpreted as the leakage rate of the \emph{D} SQUID resonator multiplied by the distance in phase space of the two coherent states corresponding to $n$ and $n+1$ photons in the \emph{N} resonator. With a detuning $\Delta=-g(n + 1)$, the value of $\Gamma_{m}$ is optimized for a given driving strength amplitude $\bar\varepsilon_D$. The driving strength can be raised to the point where the number of photons in the \emph{D} resonator reaches the maximum photon number  $n_{\mathrm{max},D}$. In this optimal condition, the measurement rate is
\be
\Gamma_m^{n,n+1}=\frac{4n_{\mathrm{max},D} g^2}{\kappa} \left [1-\exp(-t\kappa/2)\right ]^2.\label{eq:Gamma}
\ee

For a homodyne measurement, the signal to noise ratio for distinguishing $n$ from $n+1$ photons is related to the measurement rate through the relation~\cite{Gambetta_2008}
$
\t{SNR}=(\int_0^{T_{m}}\sqrt{\eta\Gamma_m(t)}\t{d}t )^2/T_{m}
$,
where  $T_{m}$ is the integration time and $\eta$ is the detection efficiency for the wave reflected off the \emph{D} resonator, in practice limited by the available amplifier. Using Eq.~\eqref{eq:Gamma}, we find
\be
\t{SNR}=\eta 4 n_{\mathrm{max},D} (g T_m)^2 \frac{[\kappa T_{m}-2(1-e^{-\kappa T_{m}/2})]^2}{(\kappa T_{m})^{3}}. \label{eq:SNR}
\ee
For a given measurement time $T_{m}$, the maximum  value of the SNR, $\t{SNR}_{\mathrm{max}}=0.32\, \eta n_{\mathrm{max},D} (g T_m)^2$, is attained when $\kappa=4.3/T_{m}$. $T_{m}$ should be chosen to be of the order of the relaxation time $T_{n+1, n}$ of the \emph{N} resonator, as further increasing $T_{d}$ will corrupt the signal to noise ratio. The optimization of the measurement amounts to maximizing $g^2 n_{\mathrm{max},D}$, based on Eq.~\eqref{eq:g}. The result of the optimization is only weakly dependent on the resonance frequencies, as long as linear coupling is effectively suppressed ($|\omega_D-\omega_N| \gg |J|$). High-efficiency microwave transistor based amplifiers with $\eta \approx 1/20$ are available over a wide frequency range in the GHz domain, which allows for flexibility in choosing $\omega_D$ given any $\omega_N$. Note: with the advent of quantum limited amplifiers~\cite{clerk_2010_QNoiseMeasAmp}, it will be possible to have $\eta \lesssim 1$. To illustrate the optimization procedure, we choose $\omega_{D}/2\pi =7$~GHz $\omega_{N}/2\pi =4$~GHz. Even though Eq.~\eqref{eq:g} indicates that when $f_D$ and $f_N$ tend towards 1/2 the value of $g$ can reach its upper limit, we choose $f_{D}=0.45$ and $f_{N}=0.45$ in order to reduce the effects due to the asymmetry of the junctions and also to avoid too small values for the maximum numbers of photons. We also assume the magnetic coupling constant to be $k=0.7$. We then proceed to optimizing $\beta_D\beta_N/n_{\mathrm{max},N} p_N p_D$, under the constraint $2\beta_\alpha \cos\pi f_\alpha \ll 1$ (we take $2\beta_\alpha \cos\pi f_\alpha$ to be less than 0.1 and fixed values of $\omega_{D}$ and $\omega_{N}$). Note: $p_N$ and $p_D$ are implicit functions of $\beta_D$, $\beta_N$, $\omega_D$, and $\omega_N$. In the optimization procedure we choose to consider values of the ratio $L_D/L_N$ between 0.5 and 2. The reason for this is that very dissimilar values of the loop inductances make it difficult to reach large values of the magnetic coupling constant $k$ with simple circuits. The optimization of the SNR over $\beta_{D}$, $\beta_{N}$, and $L_D/L_N$, leaves still one free scaling parameter, which we take to be $L_{J,N}$. We set $L_{J,N}$ to be 1.1 nH, corresponding to a critical current $I_{c,N}=300$~nA.

\begin{figure}[!]
\includegraphics[width=3.4in]{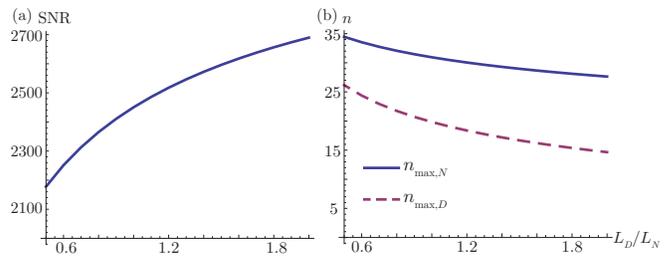}
\caption{\label{fig4}(Color online) (a) The $\t{SNR}$ as a function of $L_D/L_N$. (b) The maximum number of photons that can be store in each resonator $n_{\mathrm{max},N}$ and $n_{\mathrm{max},D}$ versus $L_D/L_N$.}
\end{figure}

In Fig.~\ref{fig4}a we show the optimal $\t{SNR}$, for $\eta=1/20$, versus the $L_D/L_N$ ratio. With a value of $T_{m}=1\, \mu$s, consistent with relaxation times of superconducting resonator with Josephson junctions \cite{sandberg_2008_tunedcavity}, the SNR is above 2000, which results in a fidelity for distinguishing $n=0$ from $n=1$ above 99\%~\cite{Gambetta_2007}.  The SNR for distinguishing $n$ from $n+1$ scales as $1/n^2$ due to the $1/n$ dependence of the relaxation time $T_{n+1,n}$. Therefore we estimate that the observation of up to 10 photons in the resonator can be done with an SNR larger than 20 (state dependent) which should result in fidelities larger than  87\%. Fig.~\ref{fig4}b shows that $n_{\mathrm{max}, N}$ is large: up to 30 photons can be held in the \emph{N} resonator, and $n_{\mathrm{max},D}$ is large enough for the linear approximation to hold. We note that to reach the values of the SNR plotted in Fig.~\ref{fig4}a, the parameters $g/2\pi$, $A_D/2\pi$, $A_N/2\pi$, and $J/2\pi$ are of the order of 10~MHz. At this value of $J$, linear coupling is negligible.

In conclusion, we presented a new method for quantum non-demolition detection of photons in a microwave resonator, based on a optimized non-linear interaction. We find that large signal to noise ratio can be obtained for up to 10 photons, assuming resonator relaxation times of the order of microseconds. The quantum non-demolition nature of the detection enables the monitoring of quantum jumps and related applications to detection of weak electromagnetic fields~\cite{caves_1980_WeakForce1}.

This work was supported by an NSERC Discovery grant. JMG was supported by CIFAR and NSERC.


\end{document}